# MITIGATING PERFORMANCE LIMITATIONS OF SINGLE BEAM-PIPE CIRCULAR e⁺e⁻ COLLIDERS

M. Koratzinos, University of Geneva, Switzerland and F. Zimmermann, CERN, Geneva, Switzerland.

*Abstract*

Renewed interest in circular $e^+e^-$ colliders has spurred designs of single beam-pipe machines, like the CEPC in China, and double beam pipe ones, such as the FCC-ee effort at CERN. Single beam-pipe designs profit from lower costs but are limited by the number of bunches that can be accommodated in the machine. We analyse these performance limitations and propose a solution that can accommodate O(1000) bunches while keeping more than 90% of the ring with a single beam pipe.

## SINGLE BEAM-PIPE LIMITATION

The CEPC collider [1] is a single beam-pipe $e^+e^-$ collider with the main emphasis on 120 GeV per beam running with possible running at 45 and 80 GeV. Bunch separation is ensured by a pretzel scheme and the maximum number of bunches is limited to 50. This very small number of bunches for a modern Higgs factory introduces luminosity limitations at 120 GeV, and severe limitations at any eventual 45 GeV running.

A machine of the size of CEPC at 120 GeV ought to be designed to be operating at the beam-beam limit and not reach the beamstrahlung limit first. The best way to reach this goal is by keeping the bunch charge low and emittances as small as possible. A large momentum acceptance also helps. Another way (and the route chosen for the CEPC) is to keep the bunches as long as possible, but this gives rise to lower instability thresholds as well as to geometric luminosity loss. According to our calculations and with reasonable assumptions for the length of the FODO cell and phase advance, we arrive at an optimal number of bunches of around 120 at 120 GeV [2]. The accommodation of this number of bunches with the pretzel scheme would be more demanding.

For an eventual running at 45 GeV the limit of 50 bunches would be inadequate, as hundreds of bunches would be needed to explore the full potential of the machine [2].

## THE 'BOWTIE' DESIGN

Without changing the basic design philosophy of the project, where the single-beam pipe approach was chosen to minimise cost, we can envisage an approach that preserves the low cost of the single beam pipe and at the same time accommodates hundreds of bunches without the use of a pretzel scheme. This approach is illustrated in Figure **1**. All arcs contain a single beam pipe but the straight sections where the experiments are located are increased in length and after the RF section a series of electrostatic separators splits the two beams sufficiently far apart transversely so that separate beam pipes and magnetic elements can be used to manipulate the electron and positron beams individually, and without any parasitic collisions. The length of the electrostatic separator section would be around 100 m on both sides of the straight section. Since now the beams travel in separate beam pipes, great flexibility about the choice of collision angle is ensured. The FCC-ee is pursuing a crab waist approach which gives excellent performance at low energies and where the crossing angle is 30 mrad.

Assuming a total length of the double beam pipe to be 2×2000m, and assuming that bunches within a train can be separated longitudinally by as little as 2 m (7 ns) then 2×1000 bunches for each species can be accommodated in the machine.

The ratio of single to double beam pipe would be ~4/52 or about 8%. Note that the cost increase would be much smaller than the above figure and actually the cost per luminosity unit would be greatly improved.

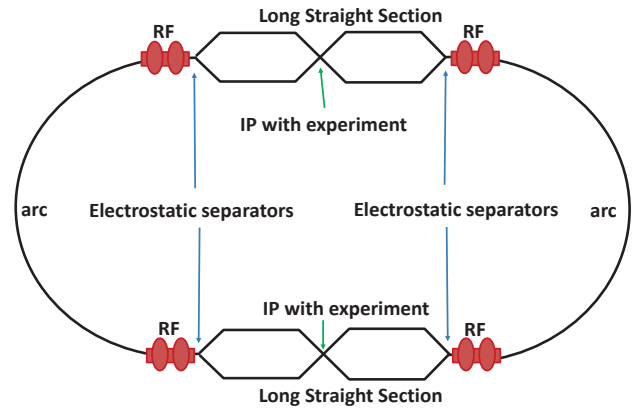

Figure 1: Schematic of the 'bowtie' idea (not to scale).

## ELECTROSTATIC SEPARATORS

For illustration purposes we have chosen the LEP electrostatic separators [3]. These were 4 m long, 11 cm wide and the maximum operating voltage was 220 kV. Each separator produced a maximum deflection of 145 µrad at 55 GeV. These separators were not without problems: aspects that need to be looked at include possible impedance issues in view of the awkward shape of the separators and reliability issues due to sparking provoked by synchrotron radiation.

Using 12 of these LEP separators (48 m total length) operated at maximum voltage would provide a separation of 20 mm ×2 to the two beams after a distance of 50 m, and a further drift space of 50 m would increase the beam separation to 60 mm ×2, a distance sufficient for the introduction of a double beam pipe and individual magnetic elements. Alternatively, a lower voltage can be used and the number of separator elements increased accordingly. Since the electrostatic separation bending radius is a lot larger than the one of the normal arc dipole magnets, synchrotron radiation issues are not expected to be a problem.

## RF SYSTEM

Bunch train operation introduces an uneven load to the RF system. All electrons and positrons arrive at the system within 2 μs for each species, whereas the RF system sees no activity for around 85 μs, i.e. the RF system is loaded only around 5% of the time. At 120 GeV per beam, where the total beam current is limited to a few mA, this does not pose a problem as with the optimum coupling factor the bandwidth of the cavities will be small compared with the revolution frequency. However 45 GeV operation at full power, where the cavities are heavily beam loaded, requires larger cavity bandwidth. Strong RF feedback will be required to counteract transient beam loading and also to damp coupled bunch modes due to the large cavity detuning needed for beam loading compensation. The problem is reduced considerably if 45GeV operation is performed at the reduced power of 20MW (20% of the nominal RF power at 120 GeV) as suggested in [2].

## ELECTRON CLOUD ISSUES

Such small separation between positron bunches can give rise to electron cloud issues. Various mitigation techniques exist including the use of amorphous carbon coating of the beam pipe [4].

## CONCLUSIONS

Single beam-pipe accelerators like the CEPC are severely limited by the maximum number of bunches that can be accommodated by a Pretzel scheme. A simple idea is presented where less than 10% of the ring is replaced by a double beam pipe and a bunch train scheme is used. Hundreds of bunches can be accommodated in such a design increasing the luminosity at 120 GeV and dramatically increasing the 45 GeV capabilities of the project.

## ACKNOWLEDGEMENTS

We gratefully acknowledge the input of A. Butterworth and J. Wenninger and the encouragement of K. Oide to publish the idea.

## REFERENCES

[1] "The CEPC pre-conceptual design report," IHEP and CAS, Beijing, 2015.

[2] M. Koratzinos, "CEPC design performance considerations," in *the 55th ICFA Advanced Beam Dynamics Workshop on High Luminosity HF2014*, Beijing, 2014.

[3] W. Kalbreier et al., "Layout, Design and Construction of teh Electrostatic Separation System of the LEP e+e- Collider," in *EPAC 88 Vol 2 pp 1184-1186*, 1988.

[4] C. Yin Vallgren et al.,, "Amorphous carbon coatings for the mitigation of electron cloud in the CERN Super Proton Synchrotron," *Phys. Rev. special topics - Accelerators and beams 14, 071001 (2011)*.